\documentclass{aa}
\usepackage{times,float}
\usepackage{epsfig}
\usepackage{graphics}
\usepackage{amssymb}

\begin{document}

\title{The search for the host galaxy of the gamma-ray burst GRB~000214 
\thanks{Based  on  observations obtained at the 
ESO 3.6 m telescope under ESO programme 165.H-0464(I).}}

\titlerunning{Searching for the GRB\,000214 host galaxy}

   \author{S.    Guziy              \inst{1,2} 
   \and J.       Gorosabel          \inst{1} 
   \and A.J.     Castro-Tirado      \inst{1} 
   \and A.       de Ugarte Postigo  \inst{1}
   \and M.       Jel\'{\i}nek       \inst{1}
   \and M.D.     P\'erez Ram\'{\i}rez \inst{3}
   \and J.M.     Castro Cer\'on     \inst{4} 
   \and S.       Klose              \inst{5}
   \and E.       Palazzi            \inst{6}
   \and K.       Wiersema           \inst{7}
   }
 
\offprints{S. Guziy}

\institute{Instituto de Astrof\'{\i}sica de Andaluc\'{\i}a (IAA-CSIC), Apartado de Correos, 3.004, E-18.080 Granada, Spain.
      \and Nikolaev State University, Nikolska 24, 54030 Nikolaev, Ukraine.
      \and Universidad de Ja\'en, Departamento de F\'{\i}sica (EPS), Virgen de la Cabeza, 2, E-23.071 Ja\'en, Spain.
       \and Astronomical Observatory, University of Copenhagen, Juliane Maries Vej 30, DK-2.100 K\o benhavn \O\ , Denmark. 
       \and Th\"uringer Landessternwarte Tautenburg, Sternwarte 5, D-07778 Tautenburg, Germany.
       \and Istituto di Astrofisica Spaziale e Fisica Cosmica, CNR, Sezione di Bologna, via Gobetti 101, I-40129 Bologna, Italy.
       \and University of Amsterdam, Kruislaan 403, 1098 SJ Amsterdam, The Netherlands.
         }
         
\mail{\tt gss@iaa.es}

\date{Received  / Accepted }

%%%%%%%%%%%%%%%%%%%%%%%%%%%%%%%%%%%%%%%%%%%%%%%%%%%%%%%%%%%%%%%%%%%%%%%%%%%%%%%%

\abstract{ We  present UBVRI-band  observations taken $\sim$300  days after
the  BeppoSAX $\gamma$-ray  burst  GRB~000214.   This GRB  did  not show  a
detectable  optical  afterglow,  however  due  to  the  localization  of  a
previously unknown,  fading, X-ray  source at a  tentative redshift  in the
range 0.37--0.47, we have searched with the ESO 3.6~m telescope for objects
with  photometric redshifts  consistent with  the mentioned  X-ray redshift
range.  We report four host galaxy  candidates, which might be the subject of
future  spectroscopic observations  in  order to  confirm their  redshifts.
 }

\maketitle

\section{Introduction}
\label{introduction}

GRB\,000214 was detected by both the  GRB monitor (GRBM) and the Wide Field
Cameras  (WFC)  on  board  BeppoSAX   on  14  February  2000,  01:01:01  UT
(\cite{Piro00}).  In the GRB monitor  it exhibited a duration of $\sim$9 s,
and a  40--700 keV fluence  of $\sim$1.4 $\times$ 10$^{-5}$  erg cm$^{-2}$.
In  the WFC  (2--30 keV)  the duration  was $\sim  $115 s  and  the fluence
$\sim$1.0  $\times$  10$^{-6}$  erg cm$^{-2}$  (\cite{Paol00}).   Follow-up
observations with the BeppoSAX Narrow-Field Instrument (NFI) began about 12
hr after  the burst.  A previously  unknown X-ray fading  point source 1SAX
J1854.4-6627, was detected in the MECS and LECS field of view at a position
of          R.A.           (J2000)=18$^{h}$54$^{m}$27.0$^{s}$,          Dec
(J2000)=-66$^{\circ}$27$^{\prime}$30$^{\prime\prime}$     (error     radius
50$^{\prime\prime}$) with a  2--10 keV flux of 5 $\times$ 10$^{-13}$ erg
cm$^{-2}$   s$^{-1}$  (Antonelli  et   al.   \cite{Ant00a}).    Within  the
50$^{\prime\prime}$ radius of the  NFI error circle, radio (\cite{Subra00})
and  IR  (Rhoads et  al.   \cite{Rhoa00})  observations  did not  find  any
variable source.  An estimation of the  redshift based on  the Fe K$\alpha$
X-ray emission  line yielded  0.37--0.47 (Antonelli et  al.  \cite{Ant00a},
\cite{Ant00b};  Kotake  \&  Nagataki  \cite{Kota01}).  

Here we  present optical observations of  the GRB\,000214 NFI  error box in
the UBVRI-bands in  order to search for objects  with photometric redshifts
in  the range  0.37--0.47,  which  could be  potential  candidates for  the
GRB\,000214 host galaxy. Throughout, we  assume a cosmology where $H_0= 65$
km s$^{-1}$ Mpc$^{-1}$, $\Omega_{\Lambda} =  0$ and $\Omega_{M} = 1$.

\begin{table*}[t]
\begin{center}
\caption{\label{table1}Journal   of   photometric   observations   of   the
  GRB\,000214 field with the 3.6~m ESO telescope. The magnitudes are given
  in the Vega system and are not corrected for Galactic reddening.}
\begin{tabular}{@{}lccccc@{}}
\hline
Date UT & Filter & Exp. Time &   Seeing         &Limiting Mag. & AB$_{off}$\\
        &        &  (sec)    &  ($\prime\prime$)&($3\sigma$) & \\
\hline
23.3766/02/2002 & B & 3 $\times$ 600 & 1.3 & 24.1 &-0.071\\
23.3989/02/2002 & R & 2 $\times$ 500 & 1.1 & 23.4 & 0.222\\
24.3537/02/2002 & I & 3 $\times$ 600 & 1.1 & 22.2 & 0.449\\
24.3726/02/2002 & V & 2 $\times$ 500 & 1.1 & 24.3 & 0.045\\
24.3909/02/2002 & U & 2 $\times$ 900 & 1.0 & 22.8 & 0.732\\
\hline
\label{table1}
\end{tabular}
\end{center}
\end{table*}

\begin{table*}
\begin{center}
\caption{\label{table2}Secondary standards in the field of GRB\,000214.}
\begin{tabular}{@{}lcccccccccc@{}}
\hline
  & R.A.(J2000)   & DEC (J2000) & U & B & V & R & I \\
\hline
A& 18:54:29.3 &$-$66:28:01 & 17.25$\pm$0.01 & 16.85$\pm$0.01 & 16.62$\pm$0.01 & 16.22$\pm$0.01 & 15.71$\pm$0.01  \\
B& 18:54:35.7 &$-$66:28:21 & 18.62$\pm$0.01 & 17.68$\pm$0.01 & 17.17$\pm$0.01 & 16.58$\pm$0.01 & 15.96$\pm$0.01  \\
C& 18:54:21.3 &$-$66:27:49 & 19.15$\pm$0.01 & 18.42$\pm$0.01 & 18.01$\pm$0.01 & 17.49$\pm$0.01 & 16.98$\pm$0.01  \\
\hline

\label{table2}
\end{tabular}
\end{center}
\end{table*}

\section{Observations}
\label{observations}

All observations  were obtained with  the 3.6~m  ESO telescope at  La Silla
(Chile).  The  CCD used was  a Loral 2048  $\times$ 2048 detector  giving a
5.4$^{\prime}$  $\times$ 5.4$^{\prime}$  field of  view.   The observations
were  carried out in  2$\times$2 binning  mode, yielding  a pixel  scale of
0.31$^{\prime\prime}$/pixel.   Table~\ref{table1}  displays  the  observing
log.  The photometry  performed to study the content of  the NFI error box,
is   based   on   aperture   photometry  carried   out   using   SExtractor
(\cite{bert00}) to study  the content of the NFI error  box.  The field was
calibrated  observing the  Landolt  star LTT  4816  (\cite{Land92}), at  an
airmass  similar to  that of  the  GRB.  Table  2 shows  the positions  and
magnitudes of  the selected  secondary standards present  in the  NFI field
(see Fig.~\ref{fig1}).

\section{Results and discussion}
\label{results}

48  objects located  closer than  $1^{\prime}$ from  the NFI  position were
detected in at  least three optical bands (out of  the five UBVRI filters).
The  magnitudes (and  upper  limits in  the  bands where  no detection  was
possible) of these objects were used  to feed the HyperZ code, yielding the
photometric  redshift,  extinction  ($A_{v}$),  galaxy  type  and  dominant
stellar   population  age   for  each   object  (see   Bolzonella   et  al.
\cite{Bolz00}  for more details  on the  HyperZ outputs).   The photometric
redshifts derived by  HyperZ for GRB host galaxies have  been tested in the
past  using  a sample  of  10  hosts  with known  spectroscopic  redshifts,
yielding excellent  results (specially for GRB host  galaxies classified as
starbursts; see  Table~2 of Christensen  et al.  \cite{Chri04a}).   For the
construction  of the  HyperZ synthetic  templates, we  assumed a  Miller \&
Scalo (\cite{Mill79})  initial mass function, and a  small Magellanic cloud
(SMC) extinction law  (Prevot et al.  \cite{Prev84}), typical  of GRB hosts
galaxies.

Table  ~\ref{table3} provides the  coordinates, magnitudes  and photometric
redshifts   for  our   four  best   candidates.   The   photometric  fluxes
corresponding to our  measurements have been obtained convolving  the 3.6~m
ESO  filter transmittances  with the  Loral  CCD, yielding  the AB  offsets
(AB$_{off}$\footnote{The       AB      offset      is       defined      as
AB$_{off}=-2.5\log(F_{\nu})-48.60-m_{vega}$,  where $F_{\nu}$  is  the flux
density measured in erg s$^{-1}$ cm$^{-2}$ Hz$^{-1}$, and $m_{vega}$ is the
magnitude in the Vega system.})  given in Table~\ref{table1}.

\begin{figure}[t] 
\begin{center}
\resizebox{8.8cm}{8.8cm}{\includegraphics{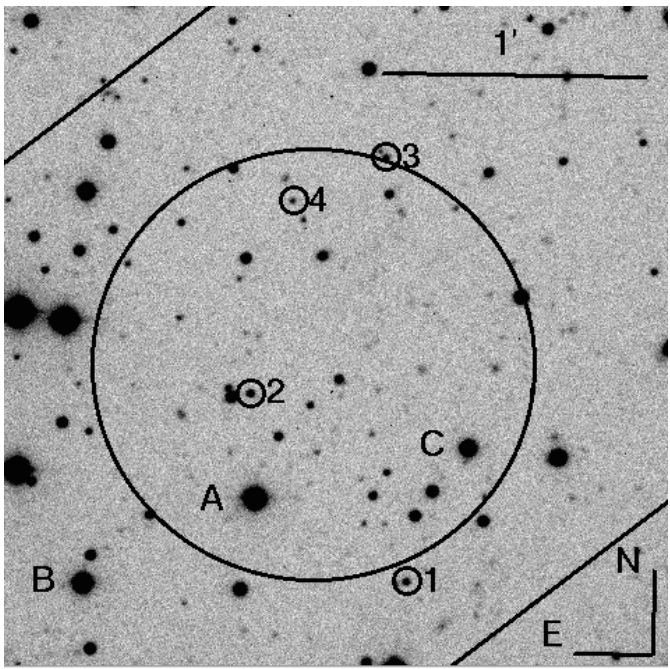}} 
\caption{V-band image  obtained at the 3.6~m ESO  telescope on 23 February
2002 of the GRB\,000214 field.  The  BeppoSAX NFI error box (solid circle) of
the  GRB\,000214 X-ray  afterglow  (Antonelli et  al.   \cite{Ant00a}) and  a
section of the  IPN annulus (area inside the solid  lines; Hurley \& Feroci
\cite{Hurl00})  are reported. The  NFI error  circle radius  is 50$^{\prime
\prime}$ with a statistical confidence level of 90\%.  The numbered objects
show  the  potential candidates  for  the  GRB\,000214  host, displayed  in
Table~\ref{table3}.  Candidate  \#1 is slightly  outside ($\sim 4.5^{\prime
\prime}$) and  \# 3  is on  the edge of  the NFI  error circle.   The stars
labeled  as  A,  B,  C  represent  the  secondary  standards  indicated  in
Table~\ref{table2}.}
\label{fig1}
\end{center}
\end{figure}

Only  object \#1  shows  a photometric  redshift  fully consistent,  within
$1\sigma$,  with  the  0.37--0.47  redshift range,  being  the  photometric
redshift of candidate  \#2 just at $1\sigma$ from  the X-ray redshift range
lower  limit.  The  two remaining  objects  (candidates \#3  and \#4)  have
photometric redshifts separated by  $2\sigma$ from the X-ray redshift range
upper limit.

However, we note that candidate  \#1 is formally outside of the 50$^{\prime
\prime}$ radius  NFI error circle and object  \#3 is just on  its edge (see
Fig.~\ref{fig1}).   Both  candidates  are  fully consistent  with  the  IPN
annulus so we decided not to discard them. Candidate \#2 is well centered in
the  NFI error  circle, but  its  photometric redshift  is only  marginally
consistent (at $1\sigma$)  with the X-ray redshift.  Thus,  inside the 90\%
confidence level NFI error box,  no object has a photometric redshift fully
consistent (within  1$\sigma$) with the  0.37--0.47 X-ray redshift  range. 

An alternative possibility is that the host galaxy of GRB\,000214 is indeed
placed within  the 50$^{\prime\prime}$ radius  NFI error circle, but  it is
fainter   in  three   or  more   filters  than   the  limits   reported  in
Table~\ref{table1}.  In this case no computation of photometric redshift is
possible and the object would be automatically discarded in our analysis. A
second alternative scenario  is possible if the GRB\,000214  host galaxy is
detected in three o more filters, but it is located in the outskirts of the
NFI error circle (i.e. on  the tail of the probability distribution).  This
might  still be  the  case for  object  \#1, which  is  located only  $\sim
4.5^{\prime \prime}$ out  of the NFI error circle  90\% boundary.  The same
conclusion stands  for object \#3  which is just  on the border of  the NFI
error circle.

\begin{figure}[t]
\begin{center}
\resizebox{\hsize}{!}{\includegraphics[bb= 80 374 544 709]{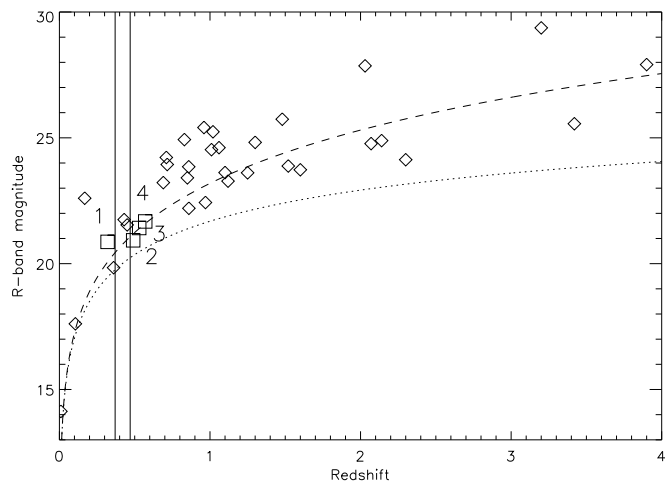}}
\caption{The  plot shows  the  R-band  magnitude of  32  GRB host  galaxies
  (rhomboids) and  the potential four  hosts of GRB\,000214 (squares)  as a
  function of redshift. All the magnitudes have been corrected for Galactic
  reddening.   The   four  candidates  have  been   labeled  following  the
  numeration  given in  Table~\ref{table3}.   The error  bars  of our  four
  candidates  are smaller  than  the  size of  their  symbols.  The  curves
  represent  the  R-band magnitude  evolution  of  a typical  $M^{\star}_R$
  galaxy  with  redshift.   The   two  vertical  lines  display  the  X-ray
  0.37--0.47  redshift bin.   The  K-corrections of  the  curves have  been
  carried out for two spectral indexes; $\beta=2$ (upper, dashed curve) and
  $\beta=0$ (lower,  dotted curve).  The  area above both  non-solid curves
  indicates qualitatively the region populated by subluminous galaxies.  As
  can be seen most of the  hosts (including the four candidates) tend to be
  subluminous  galaxies.  The  positions of  the candidates  are consistent
  with  the loci  of typical  host  galaxies at  similar redshifts  ($z\sim
  0.5$).  All  the magnitudes have  been corrected for  foreground Galactic
  extinction, using the Schlegel et al. (\cite{Schl98}) dust maps.}
\label{fig3}
\end{center}
\end{figure}

In Fig.~\ref{fig3}  we display the  R-band magnitudes compiled for  32 host
galaxies  (rhomboids) and  our  four candidates  (squares),  once they  are
corrected for  foreground Galactic extinction  (E(B$-$V)=0.061, Schlegel et
al.  \cite{Schl98}).  The curves display the apparent R-band magnitude of a
reference $M^{\star}_R$ galaxy  when it is redshifted from  $z=0$ to $z=4$.
$M^{\star}_R$ represents  the R-band absolute magnitude  and determines the
knee of  the luminosity function, separating the  intrinsically bright from
the subluminous  galaxies. We  assumed a value  of $M^{\star}_R =  -20.29 +
5~\log(H_0/100)$   (Lin  et   al.  \cite{Lin96})   estimated   adopting  an
Einstein-de Sitter Universe (as in the present study).  In order to perform
the  K-correction  (Oke  \&  Sandage  \cite{Oke68})  the  spectrum  of  the
$M^{\star}_R$  galaxy has been  assumed to  be a  power law  ($F_{\nu} \sim
\nu^{-\beta}$) with the spectral index ranging from $\beta=0$ (lower dotted
line)  to  $\beta=2$  (upper  dashed  line).   This  spectral  index  range
generates a broad set of  colours, ($0.3<$B-R$<1.2$) accounting for most of
the   galaxies  seen   in  the   Hubble   Deep  Field   (Williams  et   al.
\cite{Will96}).  As it is shown, the four candidates seem to be subluminous
galaxies, tending to be above the dotted curves.  Thus, our candidates show
apparent and  absolute magnitudes similar  to GRB host galaxies  at similar
redshifts ($z\sim0.5$).

\begin{figure*}[b]
\begin{center}
\resizebox{8.7cm}{!}{\includegraphics{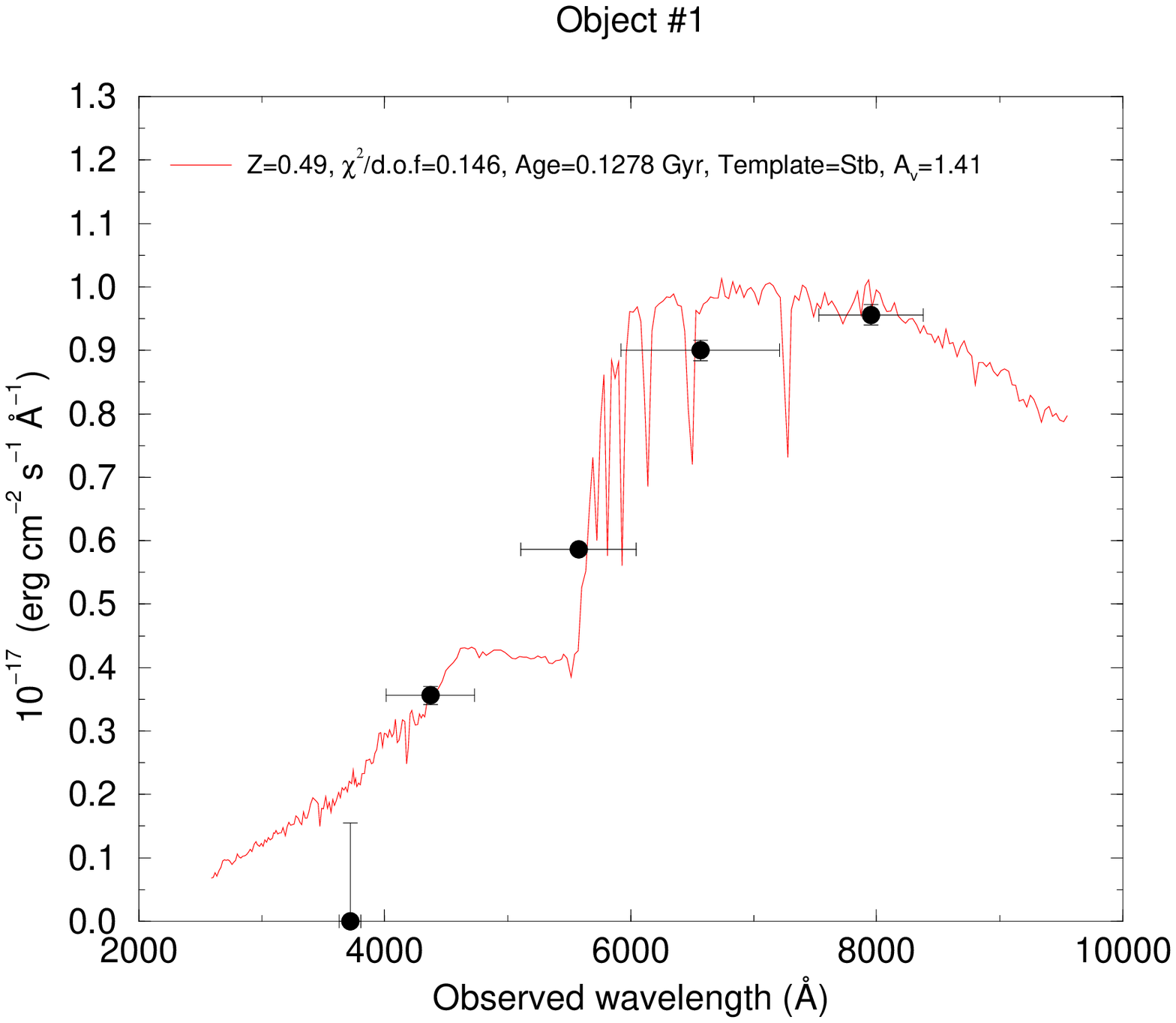}} 
\resizebox{8.7cm}{!}{\includegraphics{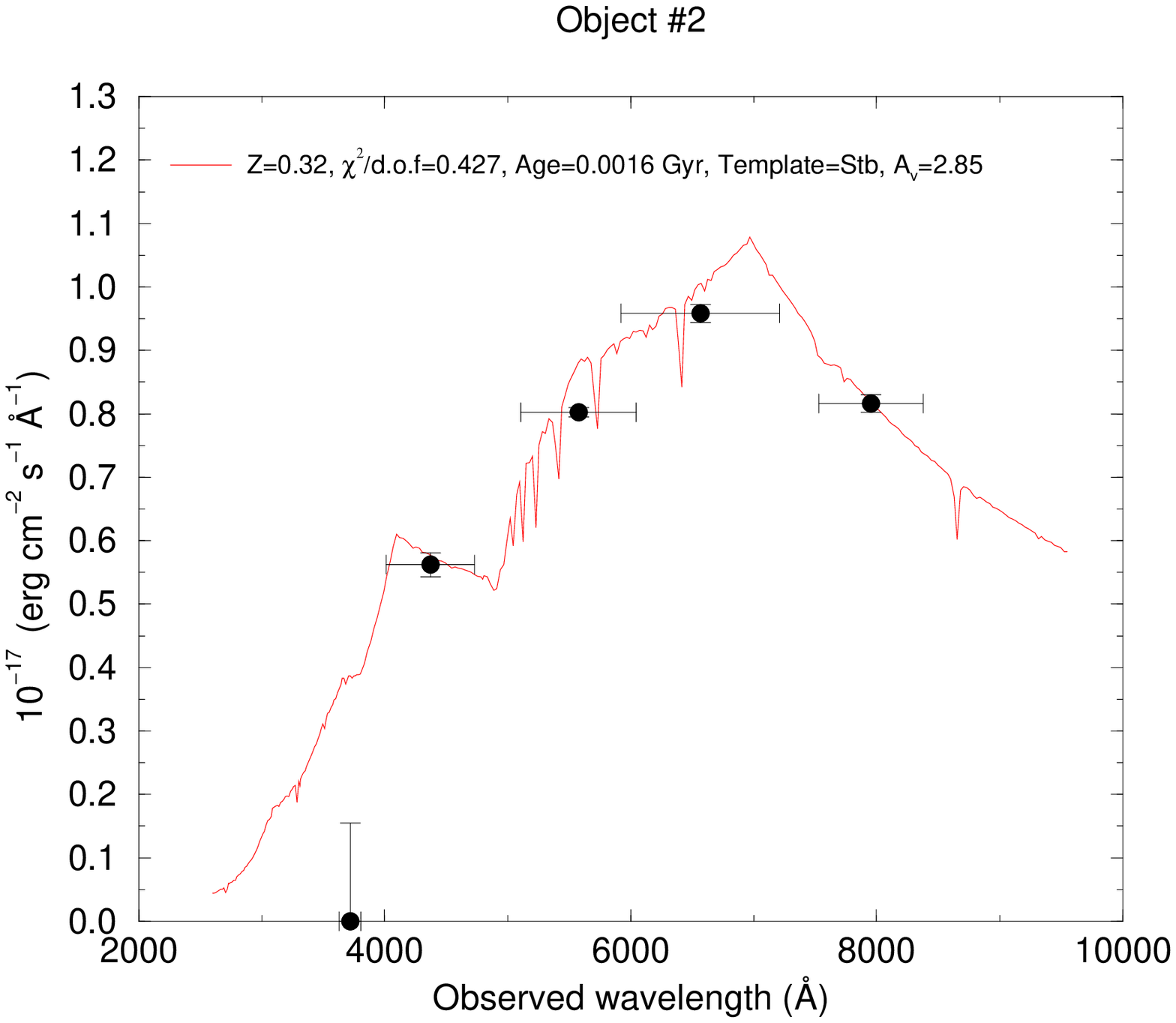}} 
\resizebox{8.7cm}{!}{\includegraphics{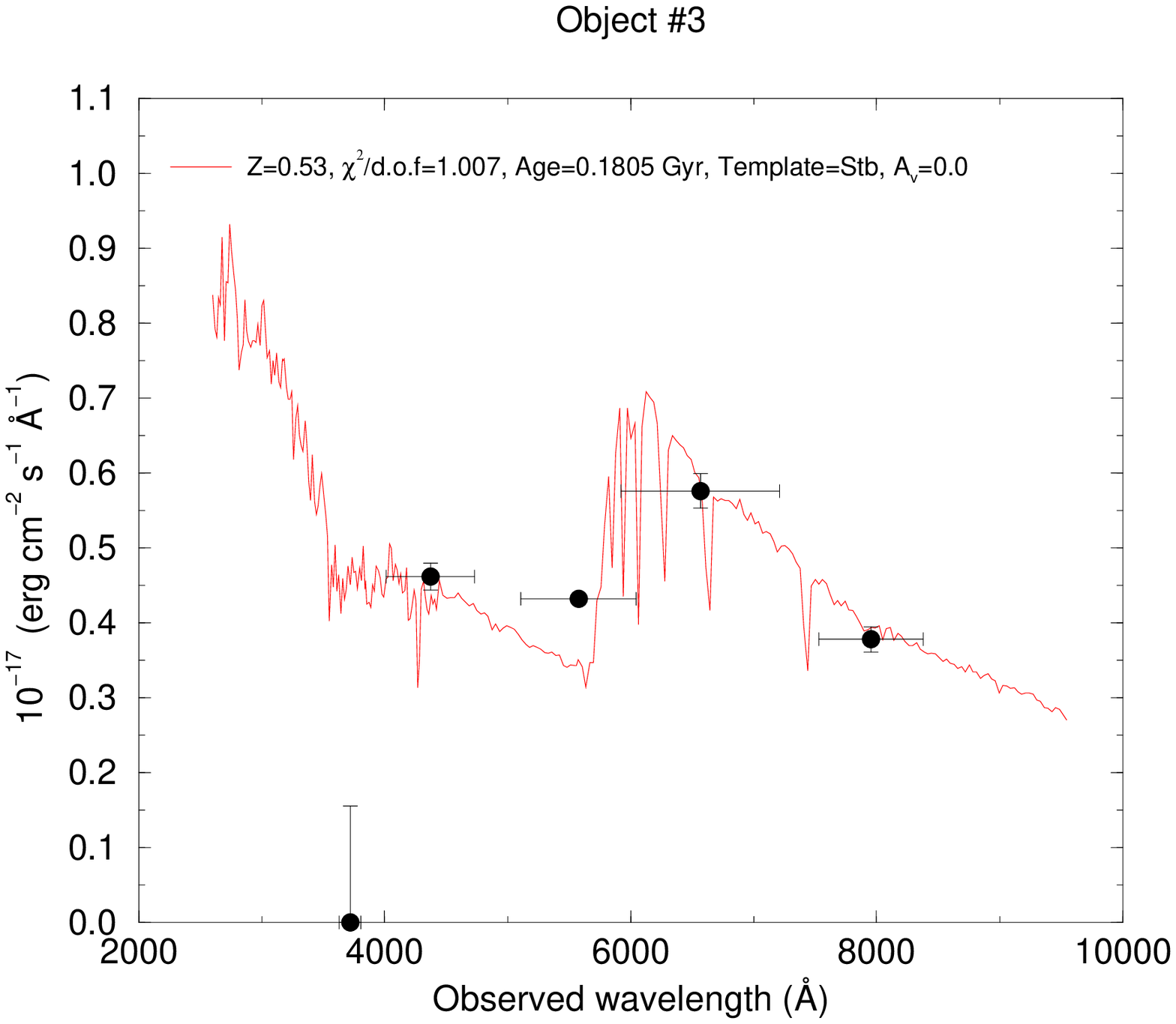}} 
\resizebox{8.7cm}{!}{\includegraphics{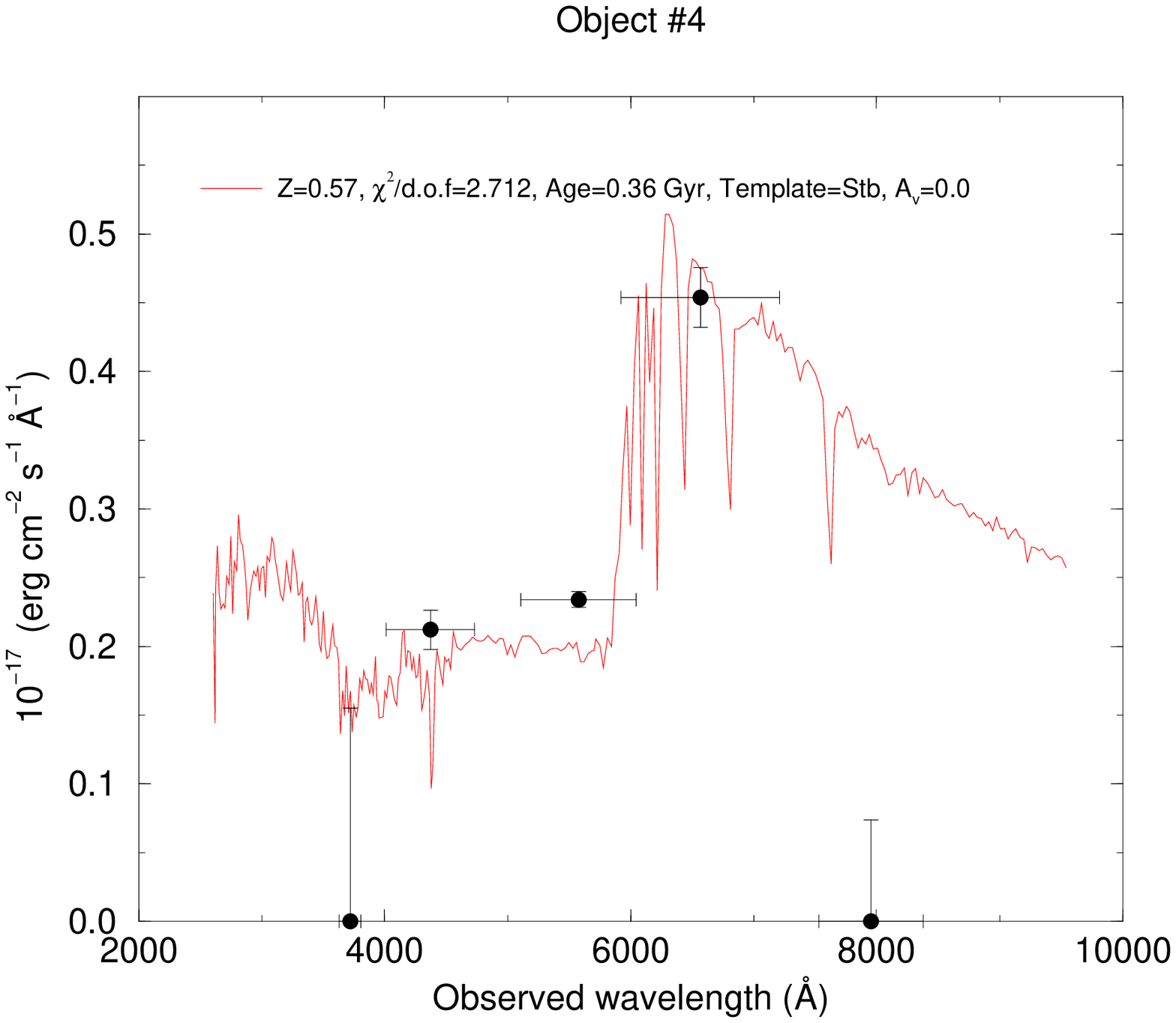}} 
\caption{Synthetic  SEDs  calculated  by  HyperZ  for  the  four  potential
candidates for the GRB\,000214 host galaxy.  The synthetic spectra are shown
by  the line and  the dots  denote the  UBVRI-band fluxes.   The horizontal
error bars  indicate the FWHM of  each filter.  The  detection upper limits
are represented by vertical error  bars ranging from the $3\sigma$ limiting
flux  (see the  associated limiting  magnitude  in the  $5^{th}$ column  of
Table~\ref{table1})  to zero  (a  more detailed  description  of the  upper
limits used by HyperZ can be found in Bolzonella et al. \cite{Bolz00}). The
inferred values  of the photometric  redshift, age of the  dominant stellar
population,  and  extinction are  given  as  inserts  in the  corresponding
panels.\label{fig2}}
\end{center}
\end{figure*}

The four objects  were classified as starbursts by  HyperZ, consistent with
the hosts' photometric spectral energy distributions (SEDs) studied to date
(Gorosabel   et   al.    \cite{Goro03a},   \cite{Goro03b},   \cite{Goro05};
Christensen  et   al.   \cite{Chri04a},  \cite{Chri04b}).    The  intrinsic
extinction  values  of  the  host  candidates range  from  $A_{\rm  v}=0.0$
(objects \#3  and \#4) to $A_{\rm  v}=2.85$ (object \#2),  while object \#1
has an intermediate $A_{\rm v}$ value of $1.41$ (see Fig. \ref{fig2}).

 Three of our four candidates  (\#1, \#2, and \#4) show compact appearance,
at least  under our seeing conditions  (see Table~\ref{table1}), displaying
full width half maxima (FWHM) similar to other ste\-llar objects present in
the GRB field.    Object \#3 is slightly extended  in the images having
the best seeing, so it corresponds  very likely to a galaxy.  The potential
stellar nature of objects \#1, \#2, and \#4 has been checked using the {\tt
CLASS\_STAR} keyword given by SExtractor.   Objects \#1, \#2, and \#4 shows
{\tt  CLASS\_STAR}   values  below  the  mode  of   the  {\tt  CLASS\_STAR}
distribution, specially in the $B$-band filter displaying {\tt CLASS\_STAR}
$<  0.8$.  Systematically  the object  with the  largest  {\tt CLASS\_STAR}
value  is  \#2.  Therefore  the  four  objects  correspond very  likely  to
galaxies, may be with the exception  of object \#2 which stellar nature can
not be completely excluded.

One potential problem might be the presence of Active Galactic Nuclei (AGN)
in our NFI error box, for  which HyperZ (not accounting for emission due to
a nebular component  or/and a central massive compact  source) would not be
an appropriate tool  to fit our SEDs. The expected  number of AGNs brighter
than z=22.5 (comparable to R=23.4, the $3\sigma$ limit of our R-band image)
closer than  $1^{\prime}$ from the NFI  position is $~\sim  1$ (Treister et
al.   \cite{Trei04}).   Thus,  for  the   sample  of  48  objects  the  AGN
contamination is expected to be only $\sim 2\%$.

\begin{table*}[b]
\begin{center}
\caption{\label{table3}  Potential  candidates   for  the  host  galaxy  of
  GRB\,000214. The table displays  the coordinates, magnitudes and inferred
  photometric redshifts for  the four best host galaxy  candidates.  All of
  them show  photometric redshifts  at or within  $2\sigma$ from  the X-ray
  redshift  range  (0.37--0.47).  Note  that  only  candidate  \#1 shows  a
  photometric redshift  fully consistent (within $1\sigma$)  with the X-ray
  redshift  bin,  although  it  is  slightly outside  the  NFI  error  box.
  Candidate \#2 is marginally (just at $1\sigma$) consistent with the GRB\,
  000214 redshift  determined from  X-ray spectroscopy. The  magnitudes are
  given in  the Vega system and  are not corrected  for Galactic reddening.
  The SEDs of these four objects can be seen in Fig.~\ref{fig2}.}
\begin{tabular}{@{}lcccccccccc@{}}
\hline
N & R.A.(J2000) & DEC (J2000)&U & B & V & R & I & photo-$z$ \\
\hline
1 & 18:54:23.34 & $-$66:28:19.8 & $>$ 22.8 & 23.34$\pm$0.11 & 22.11$\pm$0.03 & 21.09$\pm$0.04 & 20.32$\pm$0.04 & $0.49^{+0.05}_{-0.07}$ \\
2 & 18:54:29.34 & $-$66:27:36.7 & $>$ 22.8 & 23.85$\pm$0.09 & 21.77$\pm$0.02 & 21.03$\pm$0.04 & 20.49$\pm$0.04 & $0.32^{+0.05}_{-0.02}$ \\
3 & 18:54:24.38 & $-$66:26:41.4 & $>$ 22.8 & 23.06$\pm$0.11 & 22.45$\pm$0.04 & 21.58$\pm$0.11 & 21.32$\pm$0.11 & $0.53^{+0.03}_{-0.03}$ \\
4 & 18:54:27.87 & $-$66:26:51.8 & $>$ 22.8 & 23.91$\pm$0.18 & 23.11$\pm$0.06 & 21.84$\pm$0.12 & $>$ 22.2       & $0.57^{+0.06}_{-0.04}$\\
\hline
\label{table3}
\end{tabular}
\end{center}
\end{table*}

Even for objects  \#1 and \#2, which show the  highest extinction among our
four  candidates ($A_v  = 1.41$  and $A_v  = 2.85$  magnitudes),  their low
photometric  redshifts  ($z=0.49$  and   $z=0.32$)  do  not  imply  a  high
near-infrared  restframe extinction.   In particular  the K$^{\prime}$-band
limits imposed by Rhoads et  al.  (\cite{Rhoa00}) would be only affected by
intrinsic   host  extinctions   of  $A_{\rm   14400\AA}=0.3$   and  $A_{\rm
16300\AA}=0.4$,  for  objects \#1  and  \#2  respectively  (assuming a  SMC
extinction law Prevot et al.  \cite{Prev84}).  For objects \#3 and \#4, the
K$^{\prime}$-band limit is even  less extincted ($A_v \sim 0$).  Therefore,
if the  host were one of  our four objects,  then it would be  difficult to
explain  the K$^{\prime}$-band  non detection  as an  effect of  the global
intrinsic host extinction.

In fact, neither  De Pasquale et al.  (\cite{DePa03})  nor Jakobsson et al.
(\cite{Jako04})  classified  GRB\,000214  as  an  intrinsically  dark  GRB.
According  to these  authors the  K$^{\prime}$-band and  X-ray observations
reported for  this GRB are not  fast/deep enough to  constrain the physical
parameters determining the SED.  

\section{Conclusions}
\label{conclusions}

We presented  here the result of  UBVRI photometry for all  objects down to
R=23.4 inside  the GRB\,000214 error  box.  After photometric  reduction of
the images and  modeling of synthetic SEDs, we have  found no object within
the 50$^{\prime\prime}$  radius NFI error circle fully  consistent with the
redshift inferred from the X-ray spectrum.

However, we  report four host galaxy candidates  with photometric redshifts
consistent within  $2\sigma$ with the  0.37--0.47 X-ray redshift  range, so
they are still statistically acceptable.   Three of them are located inside
(or just  on the border of)  the NFI error  box, although they do  not show
photometric  redshifts   consistent  (within  $1\sigma$)   with  the  X-ray
spectroscopic  redshift  range.   A  fourth  R=21.1  mag  object,  shows  a
photometric redshift  of $z=0.49^{+0.05}_{-0.07}$, fully  consistent within
$1\sigma$.  We note  that this candidate, although consistent  with the IPN
annulus, is  slightly ($4.5^{\prime\prime}$) outside of the  90\% NFI error
circle.

We can not discard that an object fainter (in three or more bands) than our
UBVRI-band detection  limits might be  the actual GRB\,000214  host galaxy.
Further   spectrophotometric  observations  of   our  four   objects  would
definitively shed light on the reliability of the proposed candidates.

\begin{acknowledgements}

The data reported  in the present paper were taken  under the ESO programme
165.H-0464(I). We are grateful to the  ESO staff at La Silla for performing
the observations in the context  of GRACE's host galaxy programme. S.~Guziy
acknowledges the receipt  of a fellowship grant from  Spain's Ministerio de
Ciencia  y Tecnolog\'{\i}a  (ref.   SB~2003-0236), and  the hospitality  at
IAA-CSIC, where this research  was carried out.  J.  Gorosabel acknowledges
the  receipt of a  Ram\'on y  Cajal Fellowship  from Spain's  Ministerio de
Ciencia  y Tecnolog\'{\i}a.   This  research was  partially  funded by  the
Spanish ESP2002-04124-C03-01 and  AYA2004-01515 programmes (including FEDER
Funds).  We  thank N.   Masetti, E.  Pian,  and C.  Kouveliotou  for useful
conversations.   We  acknowledge our  anonymous  referee  for fruitful  and
constructive comments.

\end{acknowledgements}

\vfill
\eject

\end{document}